\documentstyle[amssym,epsfig,subfigure]{mn}

\def\lmin{\ell_{\min}}
\def\lmax{\ell_{\max}}
\def\kmin{k_{\min}}
\def\kmax{k_{\max}}
\newcommand{\bi}[1]{\mbox{\boldmath $#1$}}

\title{An adaptive filter for the PLANCK Compact Source Catalogue
construction}

\author[L.-Y.~Chiang et al.] {L.-Y. Chiang$^{1}$,  H.E.~J\o
rgensen$^{2}$, I.P.~Naselsky$^{3}$, P.D.~Naselsky$^{1,3}$,
 \and I.D.~Novikov$^{1,2,4,5}$,  and P.R.~Christensen$^{1,6}$ \\  
$^1$ Theoretical Astrophysics Center, Juliane Maries Vej 30,
DK-2100,  Copenhagen, Denmark \\ 
$^2$ Astronomical Observatory, Juliane Maries Vej 30, DK-2100,
Copenhagen, Denmark \\
$^3$ Rostov State University, Zorge 5, 344090 Rostov-Don, Russia \\
$^4$ Astro-Space Center of Lebedev Physical Institute,
Profsoyuznaya 84/32, Moscow, Russia \\
$^5$  NORDITA, Blegdamsvej 17, DK-2100, Copenhagen, Denmark \\
$^6$ Niels Bohr Institute, Blegdamsvej 17,
DK-2100 Copenhagen, Denmark}

\date{Accepted 2002 ???? ???; Received 2002 ???? ???}

\begin{document}
\maketitle

\begin{abstract}
We develop a linear algorithm for extracting extragalactic point
sources for the Compact Source Catalogue of the upcoming PLANCK
mission. This algorithm is based on a simple top-hat filter in the
harmonic domain with an adaptive filtering range which does not
require {\it a priori} knowledge of the CMB power spectrum and the
experiment parameters such as the beam size and shape nor pixel noise
level.    
\end{abstract}

\newcommand{\etal}{{et~al.~}}

\section{Introduction}

The ESA PLANCK Surveyor will produce ten all-sky high-resolution
maps of the cosmic microwave background (CMB) anisotropies at 9
frequencies, from which the angular power spectrum will be derived
to place constraints on the cosmological parameters. The maps produced
by the PLANCK satellite, however, will not only include the CMB signal but
also contain some astrophysical foreground sources arising from  dust,
free--free, synchrotron emission, Sunyaev-Zel'dovich effects and
extragalactic point sources. It is therefore an important task to
separate those foreground sources from the CMB signal.  In this paper we
will concentrate on the bright point sources with flux above $0.1-0.3$
Jy in connection with the Early Release Compact Source Catalogue (ERCSC)
from the PLANCK mission. The main goal of the ERCSC construction is to
produce such a catalogue right after the first six months of observation
of the CMB sky by the PLANCK before more complicated and time consuming
analysis of the CMB power spectrum, component separation and investigation
of in-flight systematics. Therefore, it would be very useful to
develop a fast method of the point sources extraction which needs as
little information as possible about the parameters of the experiment.

In this paper, we present a fast linear  algorithm for the extraction of
extragalactic point sources from the CMB maps, which is a
generalized amplitude-phase method of Naselsky, Novikov \& Silk (2002). There
have been developments of methods on point source extraction such as
the high-pass filter by Tegmark~and~de~Oliveira-Costa~(1998) (hereafter TO98),
Maximum--Entropy  Method (MEM) by Hobson~et~al.~(1999), and Mexican
Hat Wavelet method (MHW) by Cay{\'o}n~et~al.~(2000). Here we introduce
a simple top-hat filter for the extraction of point sources in the CMB
maps. This method works very well without assumptions about the
cosmological model, or {\it a priori} knowledge of the power spectrum
of CMB and point sources.

The filter proposed by TD98 is {\it optimal} for point source
extraction from the theoretical point of view, and requires the exact
information about the CMB and the foregrounds power spectrum, the beam
shape properties (its ellipticity and orientation), and the pixel
noise level. For concrete applications, however, other methods can
also be useful, which explains the succeeding discussions of other
filters in the literature mentioned above and in this paper. We will
compare our filter with the TD98 filter.

This paper is arranged as follows. In Section 2 we introduce the
top--hat filter and elaborate the subtlety in the definition of the
criterion by which the point sources are extracted. As our top--hat
filtering range is adaptive, we apply in Section 3 the top--hat filter
to the numerical simulated maps and estimate the filtering ranges for
the PLANCK channels when the experiment parameters such as the beam
size and the pixel noise level are known. We also compare our method
with the theoretically optimal TD98 filter in Section 4. In Section 5
we generalize our filter to an algorithm that does not need any
information about the CMB power spectrum, the beam size and the noise
level. The results and discussion are in Section 6.

\section{The top--hat filter}
In the observed map by the PLANCK satellite the signal at pixel $i$
can be expressed as 
\begin{equation}
d_i=S_i({\bi r}_i)+n_i
\end{equation}
where $n_i$ is the pixel noise, ${\bi r}_i$ correspoinds to the
position of the $i$-pixel in the map and $S_i \equiv \Delta T/T({\bi r}_i)$
includes the CMB signal and foreground contaminations, of which the
relevant component to this paper is the point source
contribution. Expanding $\Delta T/T$ in spherical harmonics, we have
\begin{equation} 
\frac{\Delta T}{T}({\bi r}_i)=\sum_{\ell m}B_{\ell m}a_{\ell m}Y_{\ell
 m}({\bi r}_i),
\end{equation}
where $B_{\ell m}$ is the beam response. The main idea of our method to 
extract point sources is through a simple linear top--hat filter in
the harmonic domain  with two cut--off scales. These two scales serve to
remove the influence of the lower and higher multipole parts of the
total power spectrum of the signal for optimal extraction of point
sources from the PLANCK maps.

In TD98 the authors introduced the ratio of the amplitude
of a point source (convolved with the beam) to the variance of the
total signal in the map $\sigma^2_{\rm tot}$, which can be denoted as 
\begin{equation}
\overline{\Re}=\left( \frac{\langle B^2 \otimes S^2
\rangle}{\sigma^2_{\rm tot}}\right)^{1/2}. 
\end{equation}
Here $\langle B^2 \otimes S^2 \rangle=\langle (\Delta T_{\rm PS}/T
)^2\rangle$ is the contribution from the point source relative flux $S$ to the
map, $\otimes$ denotes convolution, and $B$ is the beam response,
which is assumed Gaussian. They found the shape of the filter
$F_{\rm TD98}$ by maximizing  $F_{\rm TD98} \otimes \overline{\Re}$.

In our method we introduce a linear filter,  
\begin{equation}
{\cal F}_{\rm TH}(\lmin,\lmax)=\Theta(\ell-\lmin)\Theta(\lmax-\ell), 
\label{eq:heaviside}
\end{equation}
where $\Theta$ is the Heaviside step function, i.e., $\Theta(x)=0$ for
$x\leq 0$, and $\Theta(x)=1$ for $x > 0$. $(\lmin,\lmax)$ is the
filtering range of ${\cal F}_{\rm TH}$. This filter has a top--hat shape in
the harmonic domain with two characteristic scales
$\lmin$ and $\lmax$, both of which are functions of the antenna
beam shape, the power spectrum of the CMB, the power spectra of all kinds of
foregrounds and pixel noise, and possible systematic
features. When these parameters are known, we can find $\lmin$
and $\lmax$ through maximizing $F_{\rm TH} \otimes  R$ for each frequency
channel of the PLANCK mission\footnote{The values $lmin$ and
$\lmax$ can be used as the
first step of the iteration scheme introduced in
Section~\ref{iteration} to maximize the $R$ factor when the parameters
have considerable variations against predicted values.}. Here $F_{\rm TH}$
is the filter in  Eq.~(\ref{eq:heaviside}) in the real domain and $R$
is the resultant relative flux, 
\begin{equation}
R=\Re -\nu_{\min},
\label{eq:deepest}
\end{equation}
where $\Re=(B \otimes S)/\sigma_{\rm tot}$ and $-\nu_{\min}$ is defined by
$- \nu_{\min} \sigma_{\rm in}$ being the amplitude of the deepest minimum
in the map, with $\sigma_{\rm in}$ the square root of the variance of the
pre-filtered map.

Note that there is a subtle difference between the TD98 filtering and
the top-hat optimization algorithm in the definition of the criterion
by which the point sources are extracted. In order to obtain the
filter shape, the TD98 filter is defined by maximizing the
$\overline{\Re}$ ratio in their theoretical derivation whereas the
top--hat filter is defined by maximizing the $R= \Re-\nu_{\min}$
ratio. According to the prediction of bright point source
contamination in the Low Frequency Instrument (LFI) frequency range of the PLANCK mission
\cite{toffolatti}, point sources with flux above $0.1-0.5$ Jy are rare
events in a $10^\circ \times 10^\circ$ patch of the sky. For the 30
GHz frequency channel, for example, the estimated number density of
point sources is $\sim$0.3-1 source for each $10^\circ \times
10^\circ$ patch. Thus each bright point source is a peculiar peak in
the $\Delta T/T$ map and the observed amplitude of the point source in
the map is a combination of the point source contribution itself
(which is not known), the signal from the CMB plus foregrounds
convolved with beam response, and pixel noise contribution in the
pixel containing the point source signal. As is mentioned in TD98, in
order to extract point sources from the filtered map it is necessary
to introduce a criterion to screen point sources from the
`noise'\footnote{ by `noise' we mean the filtered CMB plus
foregrounds and pixel noise}. It is the so--called $5\sigma_{\rm f}$ criterion,
which means that the peaks in the filtered map with amplitudes above
$5\sigma_{\rm f}$ threshold are identified as point sources, $\sigma_{\rm f}$
being the square root of the variance of the filtered map. The amplitude of
each filtered peak above $5\sigma_{\rm f}$, however, is the combination of
the amplitude of the point source and the filtered
`noise'. Therefore, generally speaking, the final (filtered) signal
around the peak area with the amplitude around $5\sigma_{\rm f}$ is
sensitive to the actual realization of the pre-filtered `noise',
which can either increase or decrease the point source amplitude
depending on initial realization of the `noise' signal. This is the
reason why the TD98 filter can distinguish {\it mean} point source
contribution from the map. In our method for the
construction of the filter we consider the worst case, i.e., when the
point source is at the position of the deepest minimum $-\nu_{\min}
\sigma_{\rm in}$ of the signal in the map. 

\begin{figure}
\centering
\epsfxsize 4.1cm 
\subfigure[]{\epsfbox{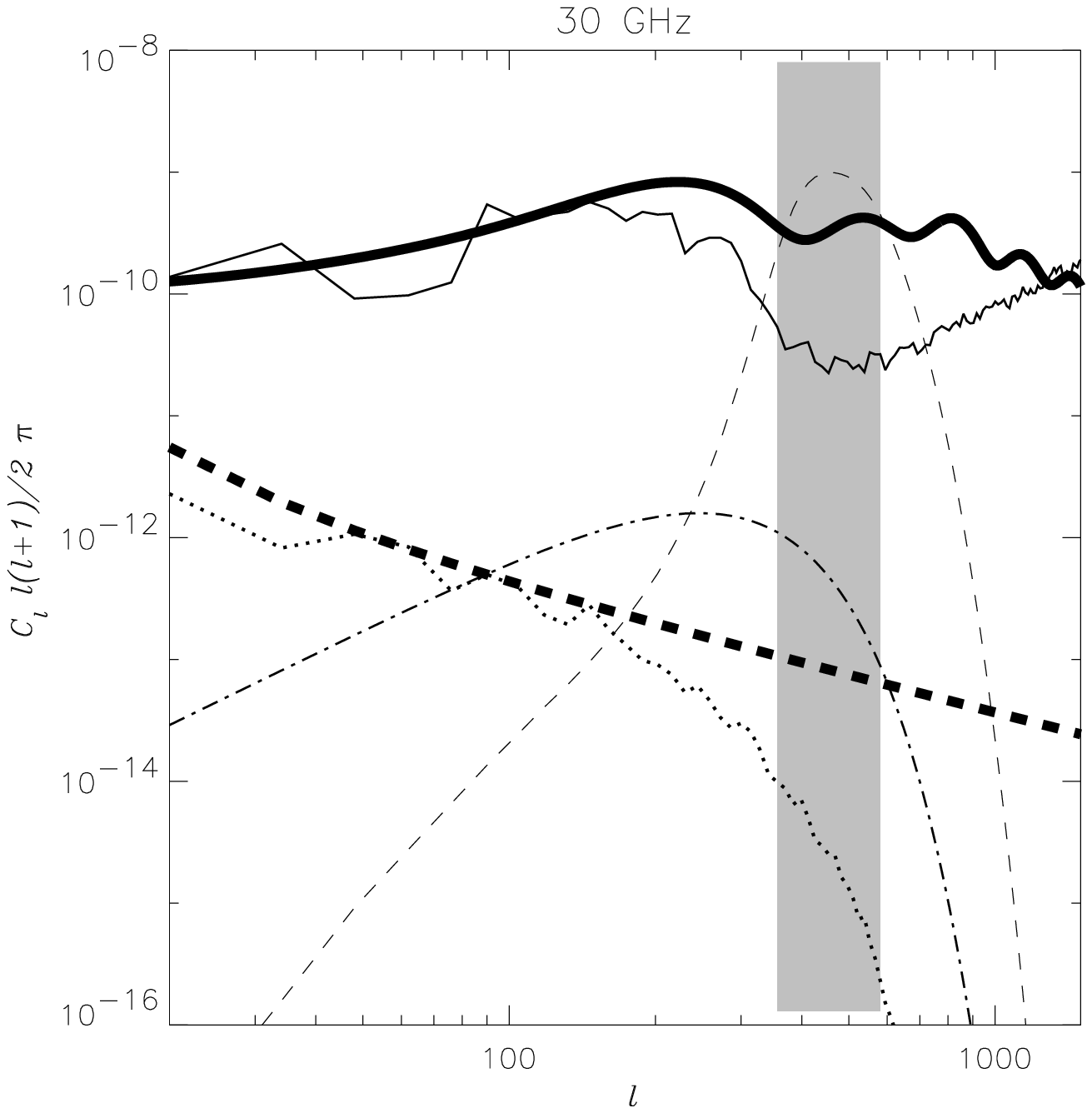}}
\subfigure[]{\epsfbox{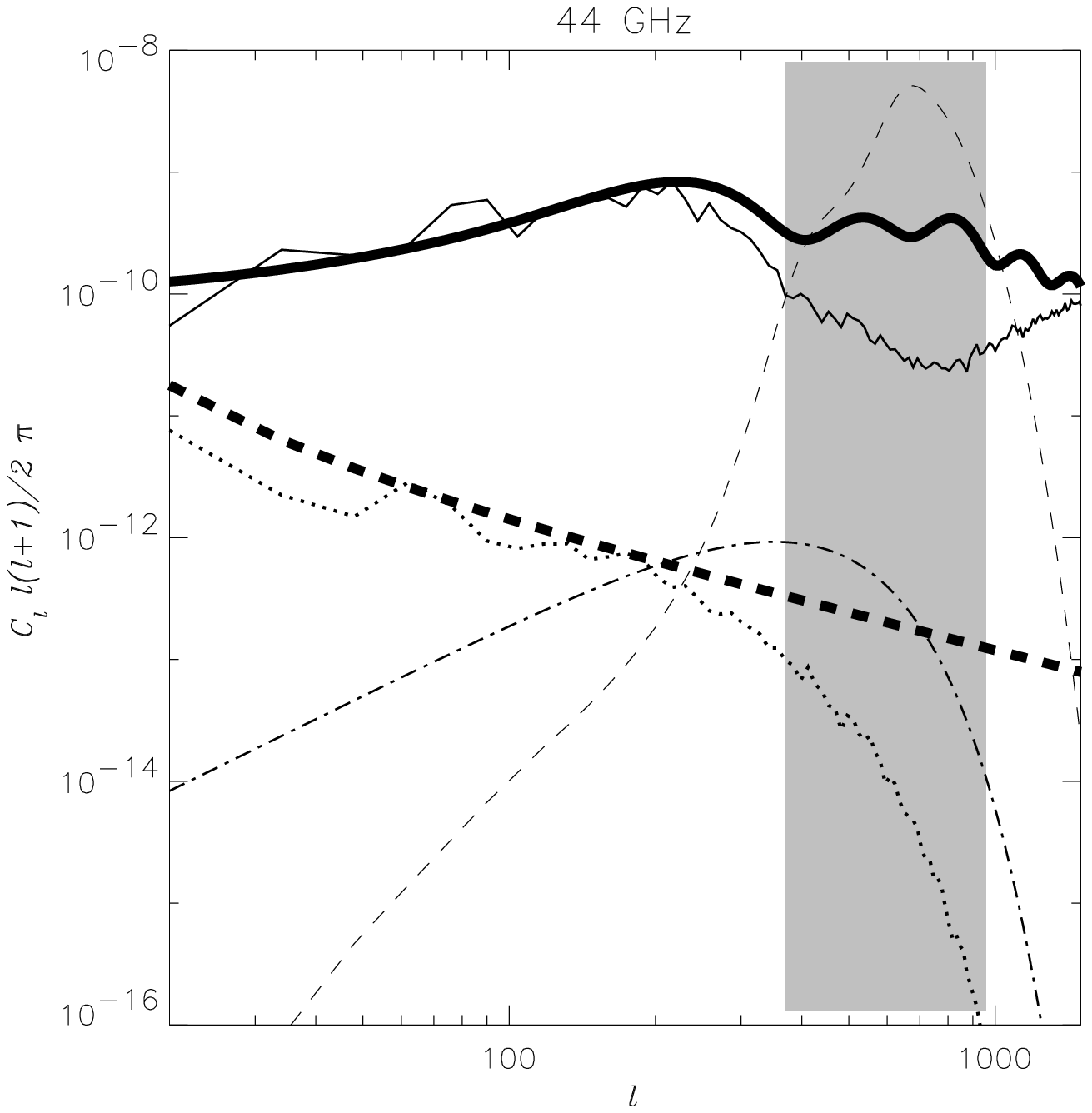}}\\
\subfigure[]{\epsfbox{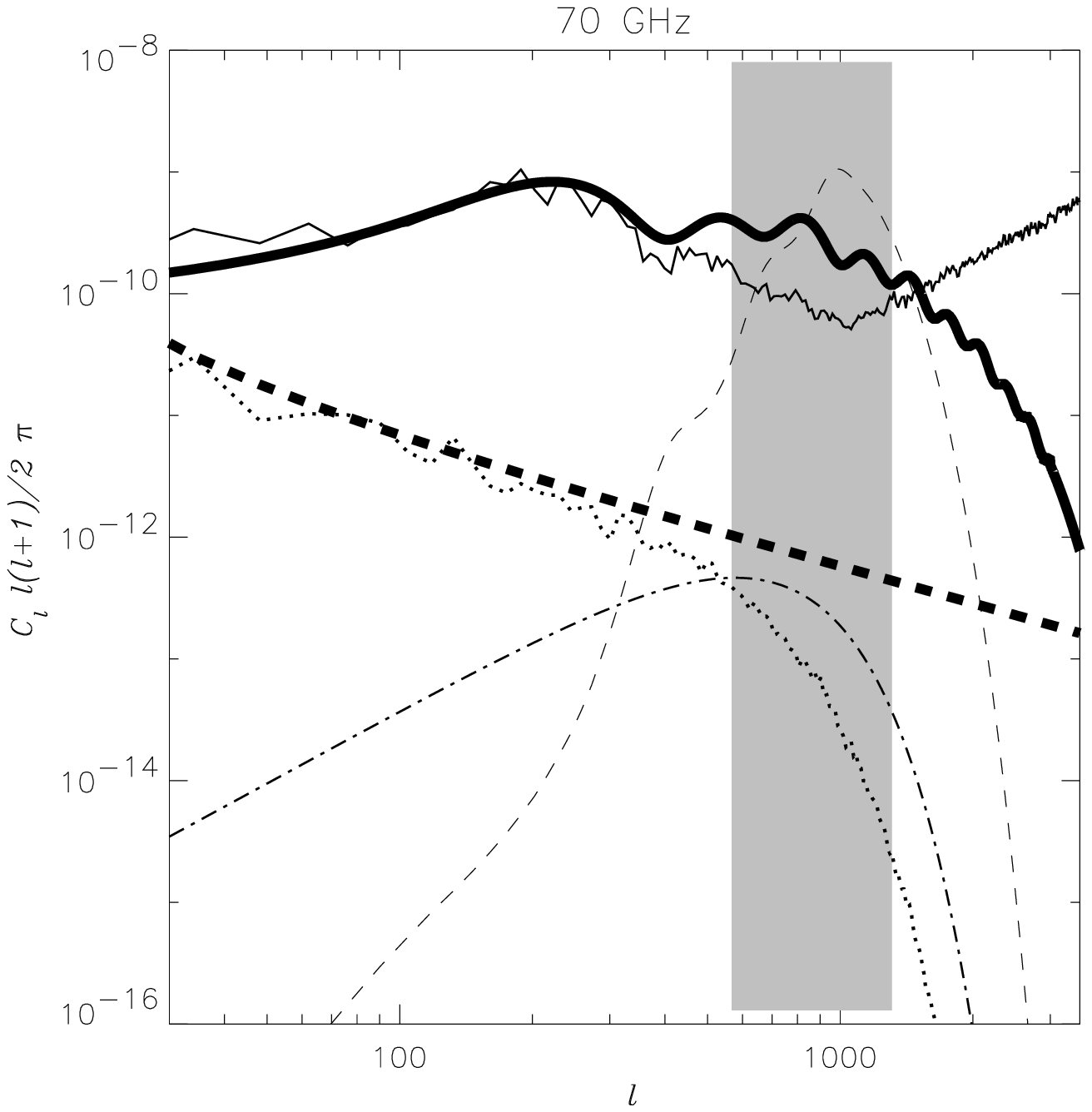}}
\subfigure[]{\epsfbox{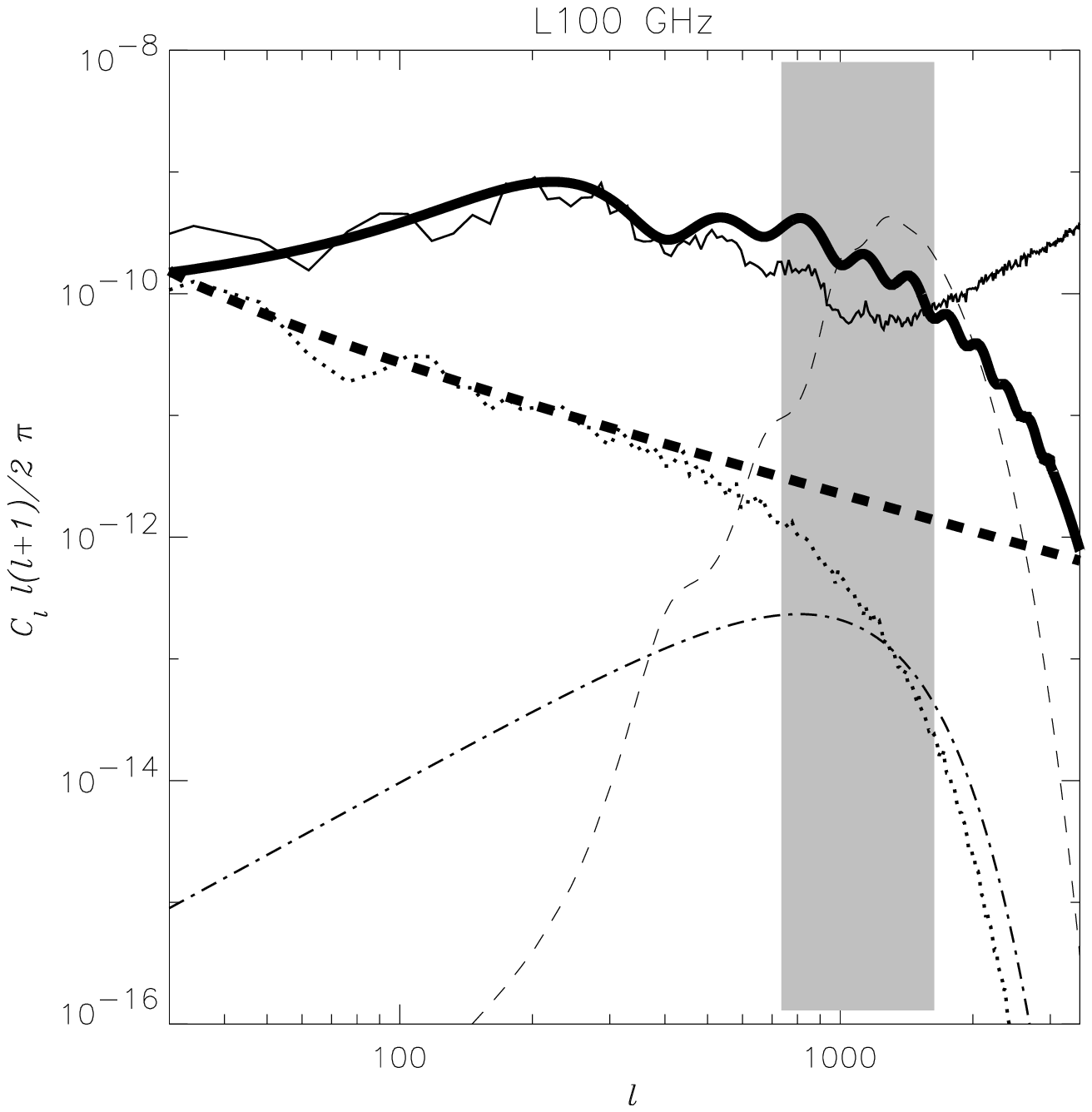}}\\
\caption{The angular power spectra and the optimal filtering range for
the LFI channels. The thick curve is the theoretical CMB power
spectrum. The thick dash line is the theoretical dust emission power spectrum,
 which is assumed $\propto \ell^{-3}$. The solid curve
represents the beam--convolved total map, which includes dust emission
(dotted line) and pixel noise. The dash--dot line is the power
spectrum of the beam--convolved point source with initial amplitude
$3\sigma_{\rm in}$, where $\sigma_{\rm in}$ is the square root of the variance of the
pre-filtered map. The shaded area shows the optimal top--hat filtering range
for the channel. For comparison, the long dash line shows the shape of
the TD98 high-pass filter, which is scaled to fit into the figure.}
\label{powerspectrum1} 
\end{figure}

\begin{figure}
\centering
\epsfxsize 4.1cm 
\subfigure[]{\epsfbox{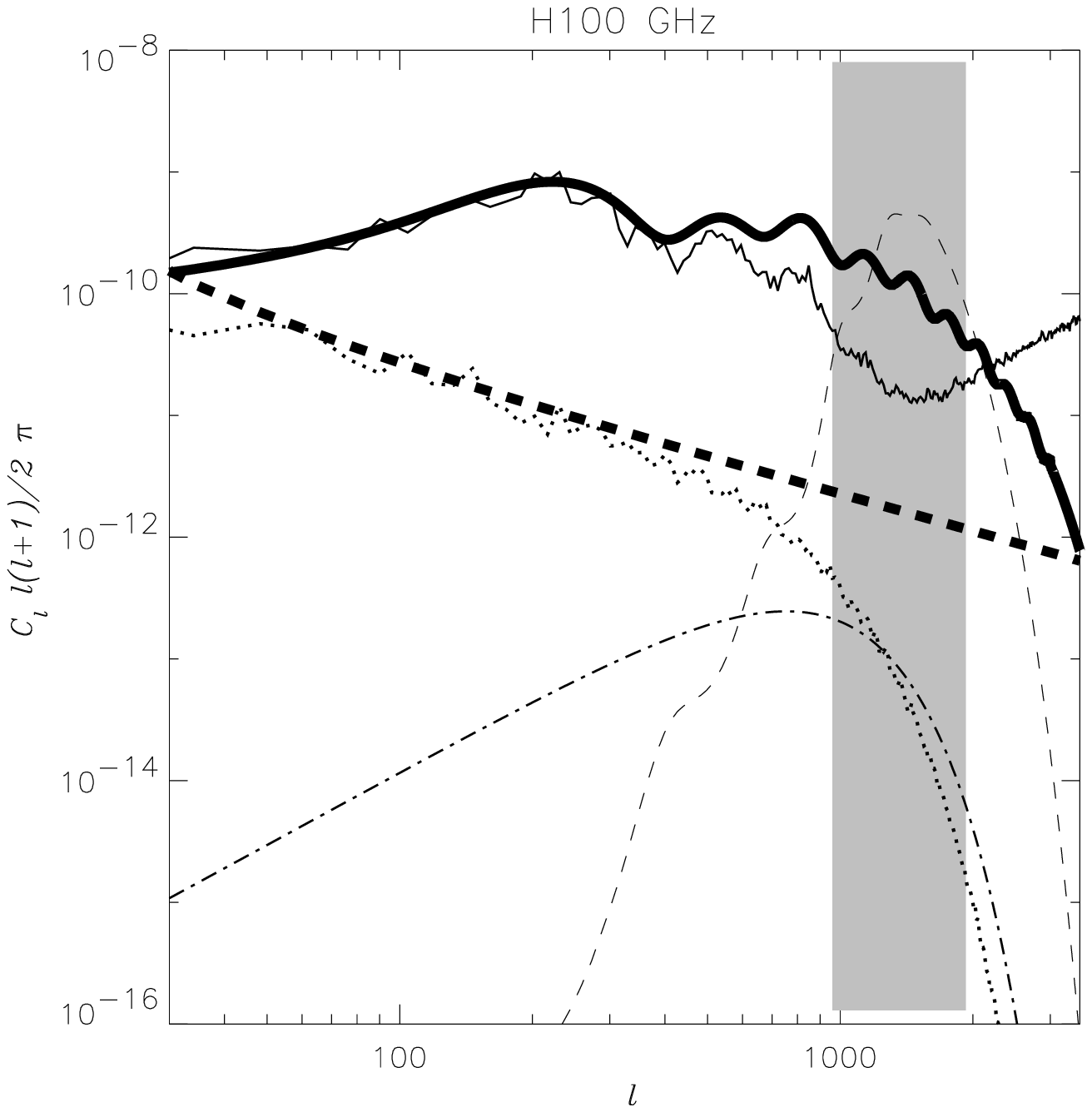}}
\subfigure[]{\epsfbox{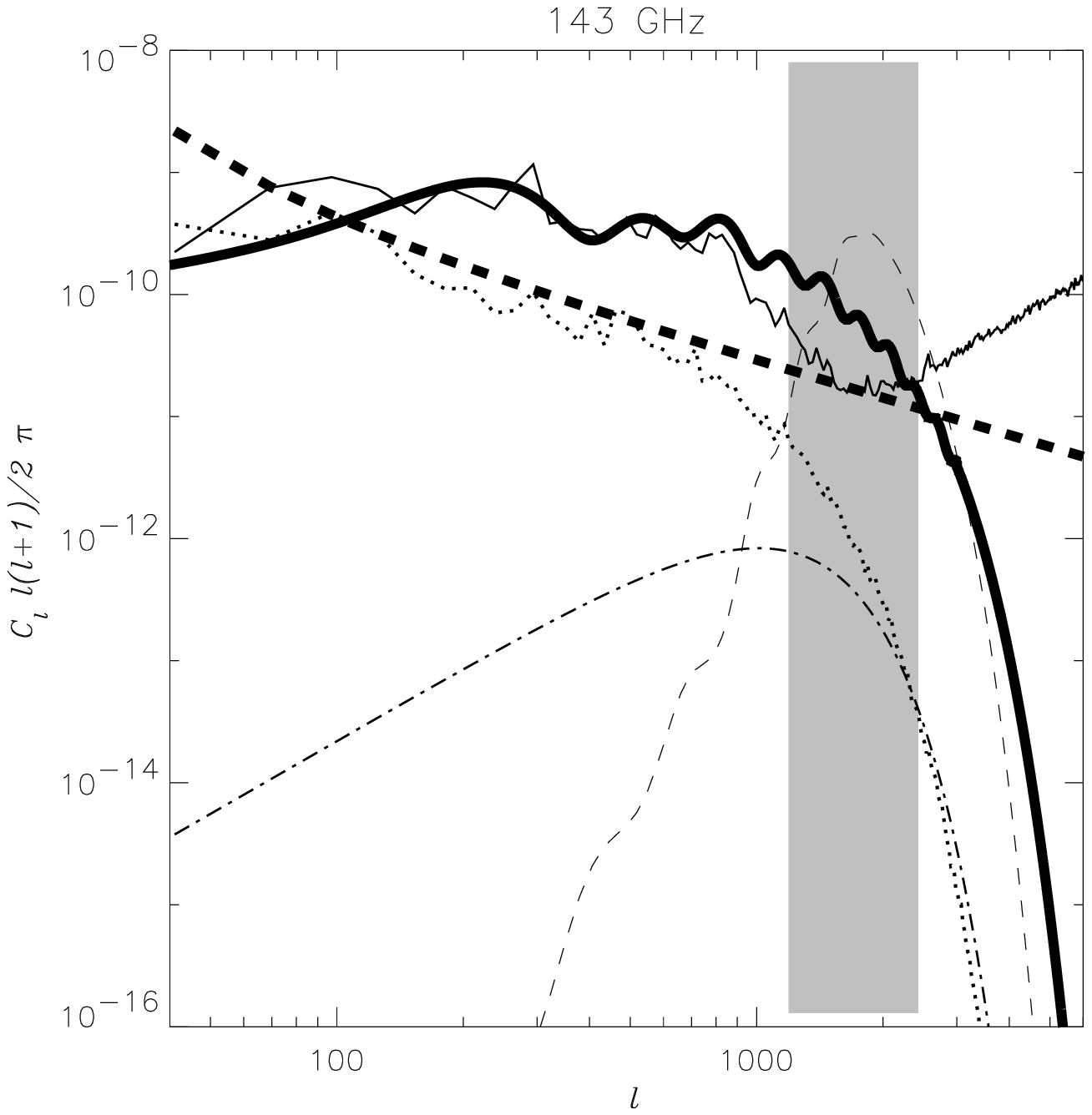}}\\
\subfigure[]{\epsfbox{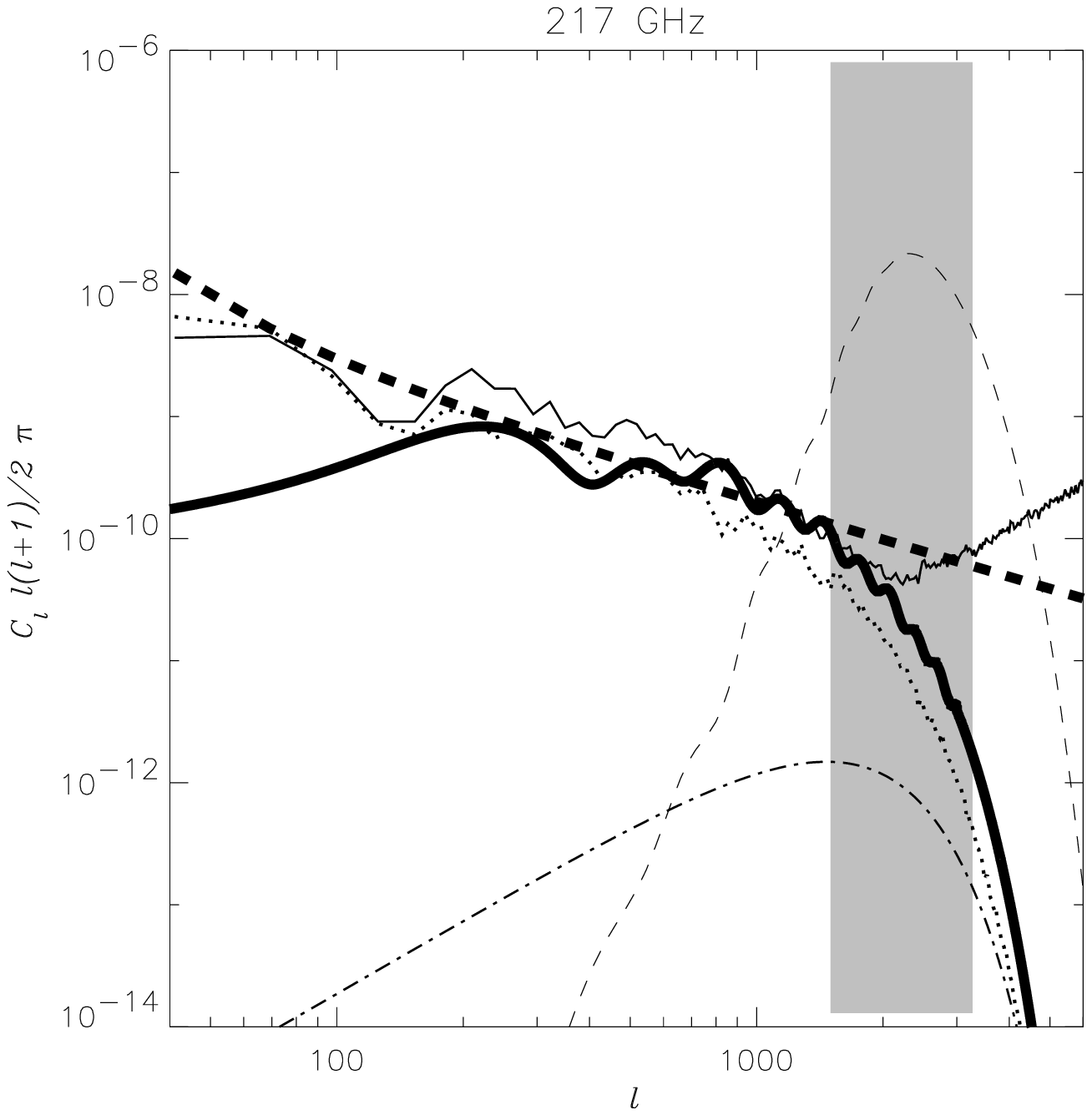}}
\subfigure[]{\epsfbox{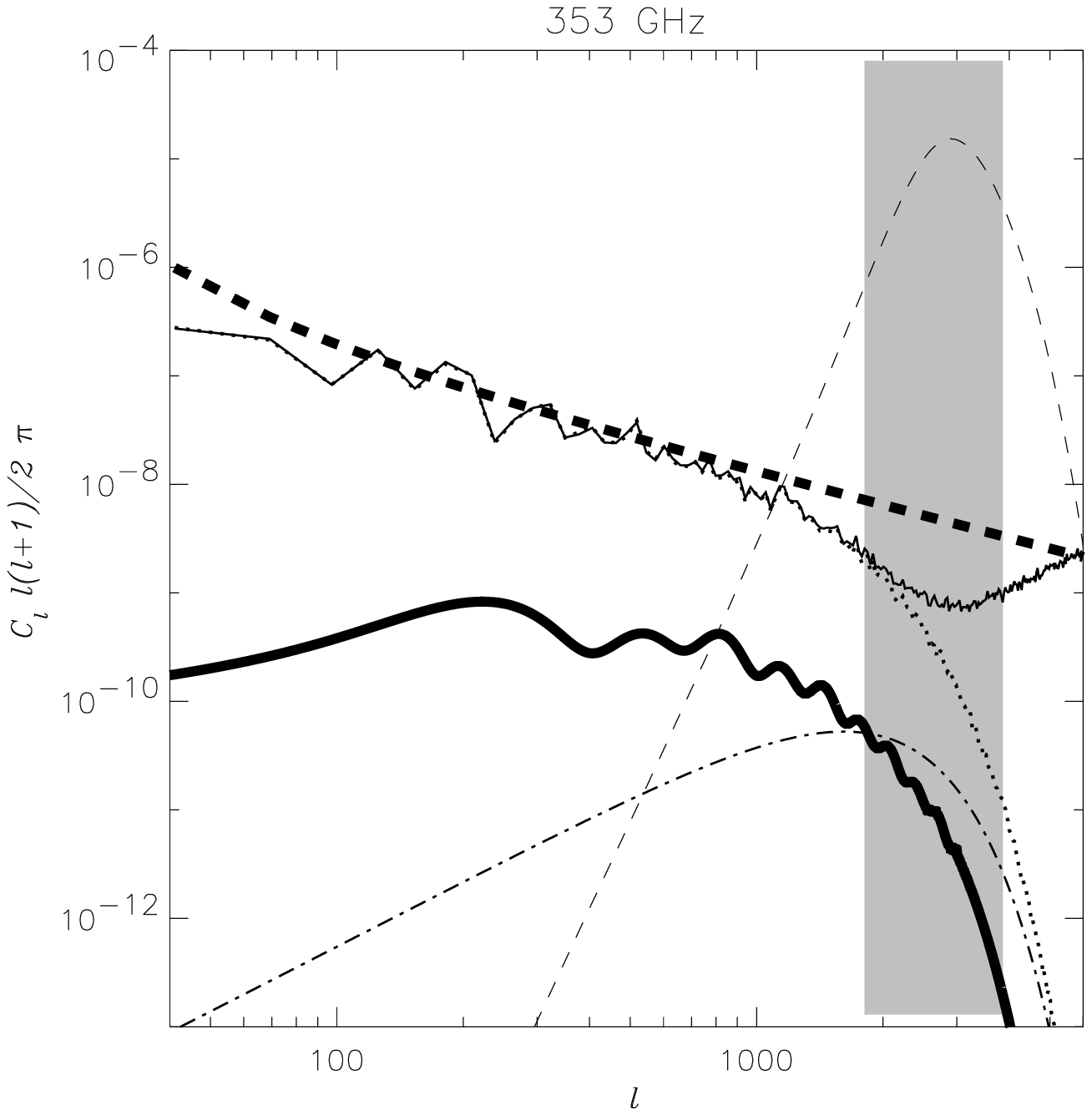}}\\
\subfigure[]{\epsfbox{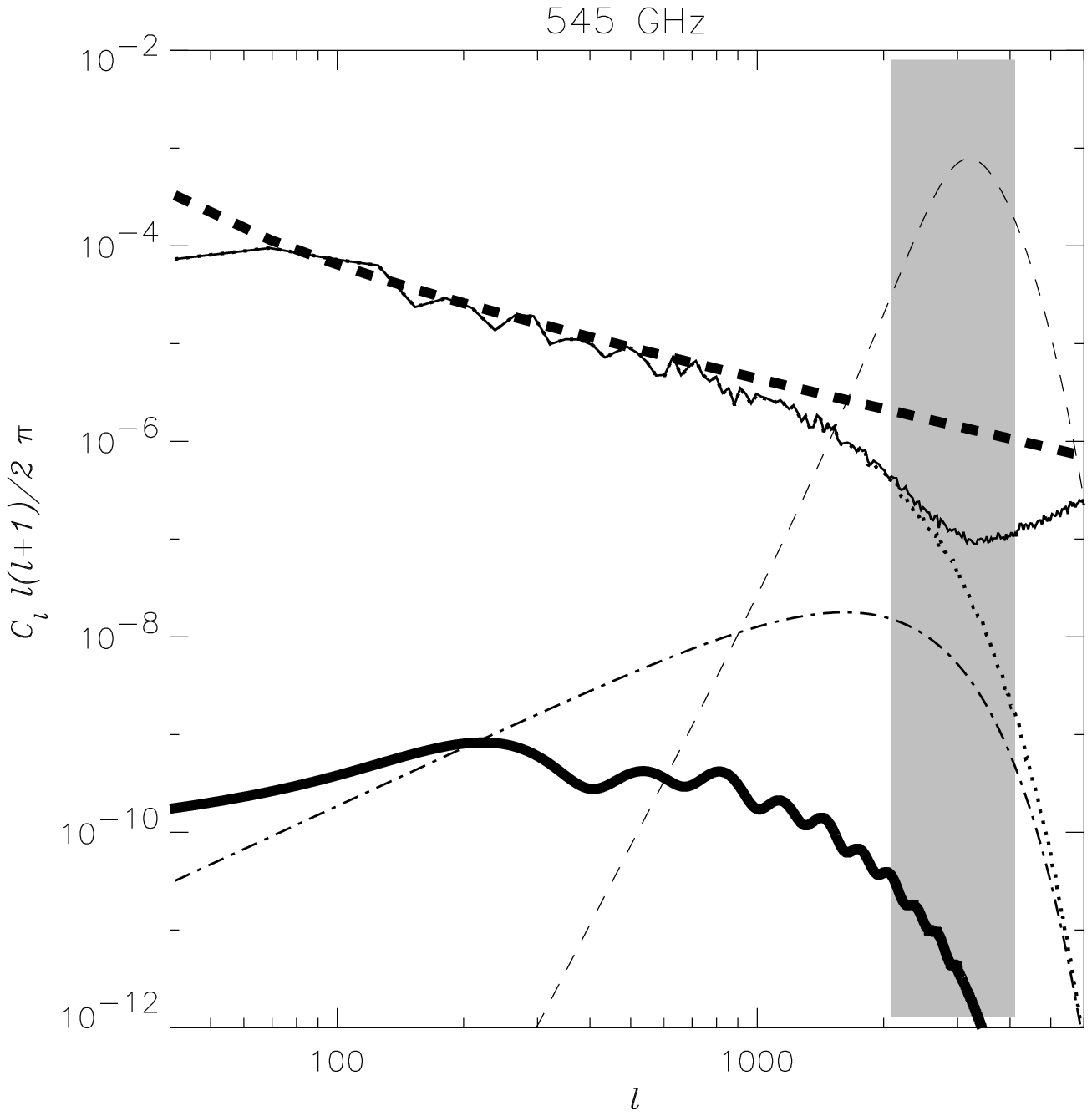}}
\subfigure[]{\epsfbox{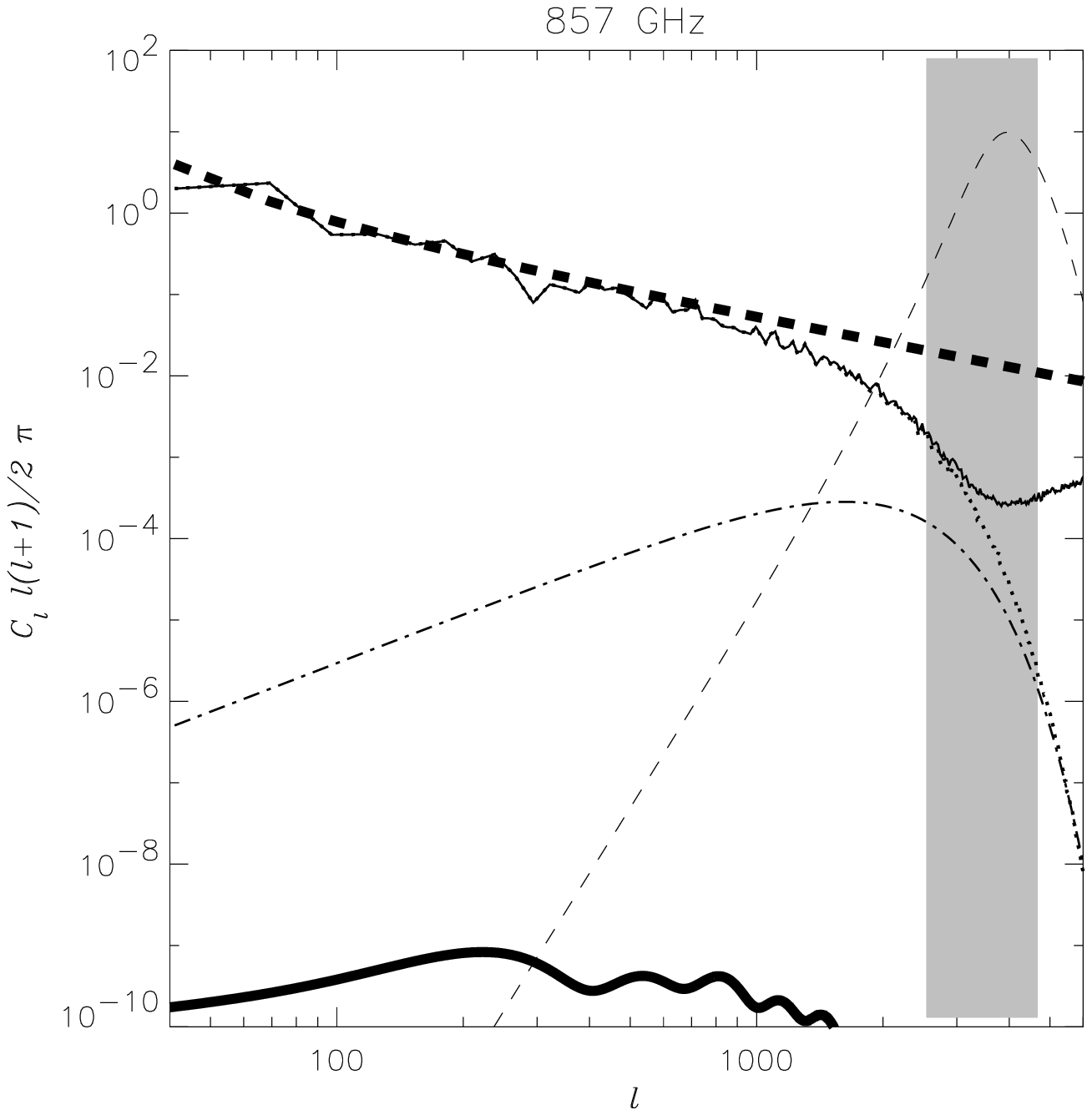}}\\
\caption{The angular power spectra and the optimal filtering range for
the HFI channels. The notations are the same as
Fig.~\ref{powerspectrum1}.} \label{powerspectrum2} 
\end{figure}

We would like to point out the significance of the
two characteristic scales $\lmin$ and $\lmax$ on enhancing the
ratio of the point source flux to the $\sigma_{\rm f}$ of the filtered
map. There are two main features on the total  power spectrum
$C_{\ell}^{\rm tot}$ from the map as shown in Fig.~\ref{powerspectrum1} and
Fig.~\ref{powerspectrum2}. On the low
multipole end, there are the CMB itself which has the standard
characteristic of $C_{\ell} \propto \ell^{-2}$ with a Harrison-Zel'dovich
power spectrum from adiabatic perturbation, together with the
so--called low multipole tail from the foregrounds such as dust
emission, free--free and synchrotron emission. Thus the scale
$\lmin$ of the top--hat filter cuts off this low
multipole part of the total power spectrum $C_{\ell}^{\rm tot}$. For the high
multipole end, on the other hand, the most dominant component in
$C_{\ell}^{\rm tot}$ is the pixel noise. The corresponding scale $\lmax$ is
therefore crucial in cutting down the pixel noise contribution in the
filtered map, hence decreasing $\sigma_{\rm f}$.

The top--hat filter aims to suppress the low multipole tail and the pixel noise
contribution so as to minimize the $\sigma_{\rm f}$. At the same time, 
it retains the part of $C_{\ell}^{\rm tot}$ that is most modified by the beam
response (see the shaded area of each panel in both
Fig.~\ref{powerspectrum1} and Fig.~\ref{powerspectrum2}). This means
that, instead of the theoretically optimal TD98 filter which needs
some preliminary and detailed information about the power spectra of
the PLANCK map components and corresponding beam response, the
top--hat filter has the simplest shape that transforms the
uncertainties of the parameters to the scales of the cut-off
$(\lmin,\lmax)$.

In constructing the top-hat filter for a specific frequency channel,
we have to estimate the two cut--off scales when the experimental
parameters such as the beam size and the pixel noise level are known. To do
so, we choose the deepest minimum of the `noise' as the position of
the point source (hence the expression of Eq.~(\ref{eq:deepest})) and
optimize the cut-off scales $(\lmin,\lmax)$ exactly from such
worst realization of the point source signal and
`noise'. The filter with the filtering range obtained by the
optimization of the signal $S_{\rm f}$ (point source) to the `noise'
$(N_{\rm f})$ ratio for the filtered map in the worst case under the
condition $S_{\rm f}/N_{\rm f}\rightarrow \max$ allows us to detect all point
sources with the same flux (and above) at any other different
locations of the map. \footnote{The deepest minimum at the initial map
as a rule is quite isolated (see Zabotin~\&~Naselsky 1985,
Bond \& Efstathiou~1987, Coles \& Barrow~1987). The
probability of finding such a realization of the point source and
`noise' is negligibly small. Thus the description `different
locations of the map' means different (and most probable) realization
of the `noise' at the point source area.}

\begin{table*}
   \begin{center}
         \begin{tabular}{|c|c|c|c|c|c|c|}
	 \hline
	 Frequency     & $\sigma_{\rm CMB}$  & $\sigma_{\rm dust}$
&
      $\sigma_{\rm noise}$ & FWHM     & Pixel size  & Simulation size \\
	 (GHz)         & ($10^{-5}$)   & ( $10^{-5}$) &   $(10^{-5})$
      & (arcmin) & (arcmin)     &  (squared area)   \\ \hline
857      & 4.47 &155700. & 2221.11& 5.0  & 1.5   & $12.8^\circ$ \\ \hline
545      & 4.47 & 1220.0 & 48.951 & 5.0  & 1.5   & $12.8^\circ$ \\ \hline
353      & 4.48 & 65.1   & 4.795  & 5.0  & 1.5   & $12.8^\circ$ \\ \hline
217      & 4.43 & 8.52   & 1.578  & 5.5  & 1.5   & $12.8^\circ$ \\ \hline
143      & 4.27 & 2.55   & 1.066  & 8.0  & 1.5   & $12.8^\circ$ \\ \hline
100 (HFI)& 4.07 & 1.15   & 0.607  & 10.7 & 3.0   & $25.6^\circ$ \\ \hline
100 (LFI)& 4.10 & 1.13   & 1.432  & 10.0 & 3.0   & $25.6^\circ$ \\ \hline
70       & 3.88 & 0.558  & 1.681  & 14.0 & 3.0   & $25.6^\circ$ \\ \hline
44       & 3.43 & 0.228  & 0.679  & 23.0 & 6.0   & $25.6^\circ$ \\ \hline
30       & 3.03 & 0.114  & 0.880  & 33.0 & 6.0   & $25.6^\circ$ \\ \hline
      \end{tabular}
\caption{We produce simulated maps (CMB signals plus dust emission
convolved with antenna beams plus pixel noise) by using the
experimental constraints at the 10 PLANCK channels by
Vielva~et~al.~(2001). Gaussian symmetric beam shape is assumed
for both the HFI and LFI channels. The 2nd, 3rd and 4th column are the
{\it rms} of the CMB, dust emission, and pixel noise, respectively (in
$\Delta T/T$).}  \label{simulation}
\end{center}
\end{table*}

\section{Estimation of the filtering range }\label{estimation}
In this section we describe the technique of the filtering range
$(\lmin,\lmax)$ estimation for the 10 maps of the
PLANCK mission. The basic model of the PLANCK experiment
which details the scan strategy, pixel noise properties, beam shape
analysis and different kinds of the foreground contaminations is
recently discussed in Mandolesi~et~al.~(2000), Burigana~et~al.~(1998),
Bersanelli et al. (1997), and Chiang~et~al.~(2001). In this section in
order to determine the filtering range we will first assume that all
the above-mentioned characteristics of the possible signals are well
determined with corresponding accuracy. This condition will be relaxed
in Section~\ref{iteration} when we introduce a more generalized
algorithm for the top--hat filtering. At this stage those
well--determined characteristics allow us to estimate the optimal
values of the $(\lmin,\lmax)$ range for the top--hat filter
(Eq.(~\ref{eq:heaviside})) and compare the efficiency of the point
source extraction from the maps by applying the TD98 and the top--hat
filter. Below we will use the flat sky approximation for the CMB maps
without loss of generality. Moreover, Chiang~et~al.~(2001) have shown
the importance of periodic boundary condition of simulations, which is
the standard part of the flat sky approximation, for descriptions of
the real signal from small patches of the sky.

\begin{figure}
\centering
\epsfxsize 6cm 
\subfigure[]{\epsfbox{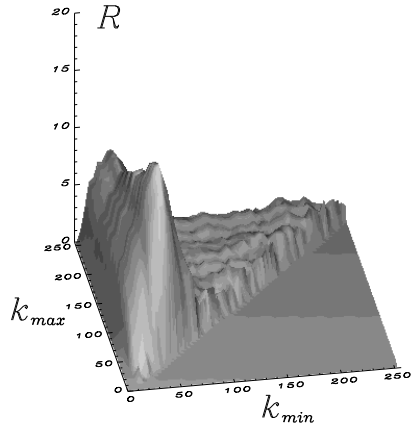}}
\epsfxsize 5.5cm
\subfigure[]{\epsfbox{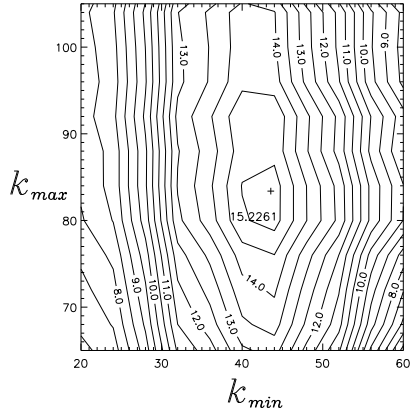}}
\caption{The display of the index $R$ as a function of $\kmax$ and
$\kmin$ for the filtered map in searching for the optimal filtering
range for the HFI 143 GHz channel. The amplitude of the point source is set
$\Re=3.21$ and located at the deepest minimum of the initial map. The
top panel shows the overall values of $R$ at different filtering
ranges. The bottom panel is the contour map near the maximum. The plus
sign marks the maximal $R$ and the contour line at 15.0 around the
maximum covers roughly 10\% of both $\kmin$ and $\kmax$. Note that
the maximum in this case does not coincide with the optimal value we
choose. Please see the context for details.} \label{contour}
\end{figure}

We produce for each PLANCK observing frequency channel a set of
realizations of simulated maps using the data provided by
Vielva~et~al.~(2001). The details of the simulations are listed in
Table~\ref{simulation}. The CMB signals are created from the angular
power spectrum of the $\Lambda$CDM model by Lee~et~al.~(2001). Dust emission
is simulated with power law index $-3$. The free-free and synchrotron
emissions are not simulated and added to obtain the filtering range
$\ell_{\min}$ and $\ell_{\max}$ in Table~\ref{range}. Without adding these two
parts of foreground emissions would, of course, affect the estimation
of the filtering range, especially for LFI frequency channels, where
the {\it rms} of both emissions are less than by one order of
magnitude or comparable to that of CMB. As the free-free and
synchrotron would have been assumed Gaussian, what would be modified
is not the filtering range but the enhancement factor $R$. Moreover,
as will become clear in Section~\ref{iteration}, these filtering
ranges serve as the initial values for the iteration scheme when we
generalize the top-hat filter. The combined realization is then
convolved with the corresponding antenna beam size. In this section we
assume Gaussian beams $B_k=\exp(-k^2\theta^2/2)$, where
$\theta=\mbox{FWHM}/2.355$, but a more realistic PLANCK antenna beam
shape can be modelled using the method proposed by
Chiang~et~al.~(2001), which, as will be shown in
Section~\ref{iteration}, can also be tackled without difficulty.

\begin{figure}

\centering
\epsfxsize 4.5cm 
\subfigure[]{\epsfbox{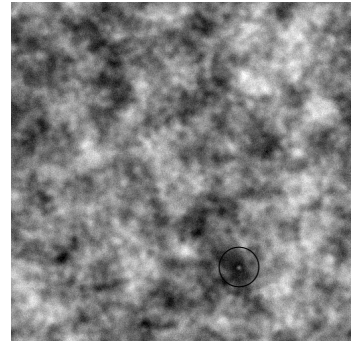}}
\put(-42,28){\circle{15}} \\
\epsfxsize 5.8cm
\subfigure[]{\epsfbox{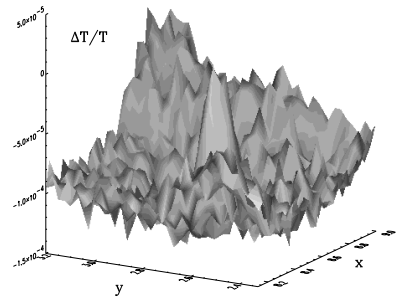}}\\
\epsfxsize 5.8cm
\subfigure[]{\epsfbox{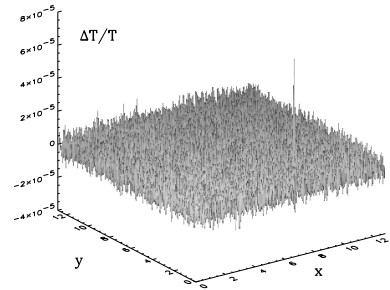}}\\

\caption{An illustration of the point source enhancing
capability. Panel (a) shows the simulated map of HFI 143 GHz with one
added beam--convolved point source of amplitude $3.21\sigma_{\rm in}$
($\Re=3.21$). The position of the point source is circled at the lower-right
quarter. Panel (b) shows the enlarged part ($x$ and $y$ axis in
degrees) of beam--convolved point source of the realization which is
located at the deepest minimum. We choose this frequency channel for
presentation because of the small beam size and the low pixel noise
level relative to the CMB signal. The shape of the beam-convolved
point source is not deformed by pixel noise due to the small FWHM and
the level of pixel noise. Panel (c) is the filtered map with the
optimal filtering range ($\kmin,\kmax$)=(42,86); the enhancement is
$R=15.50$.}
\label{illustration}
\end{figure}

To estimate the $(\kmin,\kmax)$ filtering range, in each
realization we add one beam--convolved point source with fixed
amplitude $\Re=3$, which is deliberately located at the deepest
minimum of the realization with beam--convolved CMB signal including
foregrounds plus pixel noise. In the flat sky approximation we
firstly Fourier transform the total map, then we impose the top--hat
filter as in Eq.~(\ref{eq:heaviside}) with a cut--off in Fourier domain, 
${\cal F} (\kmin,\kmax) = \Theta(| {\bf k}|-\kmin)
\Theta(\kmax - |{\bf k}|)$. We inverse Fourier transform the
filtered Fourier ring and calculate the ratio of the peak amplitude to
the $\sigma_{f}$ from the filtered map. The filtering range
$(\kmin,\kmax)$ is determined as the one from which the maximal
enhancement $R$ can be reached for the filtered map.

Figure~\ref{contour} shows the surface $R$ as functions of $\kmin$
and $\kmax$ for the High Frequency Instrument (HFI) 143 GHz
channel. The surface $R(\kmin,\kmax)$ for each frequency channel is
morphologically similar, but the position for maximal $R$ can vary
slightly from realization to realization owing to different lowest
minimal values $-\nu_{\min}\sigma_{\rm in}$ of Eq.~(\ref{eq:deepest}) and
the effect from cosmic variance. From the contour map we can easily
see that, due to the flatness around the maximum, the filtering range
covers roughly 10 per cent of $\kmin$ and $\kmax$ with only a few
percent of variation in enhancement. We therefore optimize the
filtering range from a set of realizations for each channel. In
Table~\ref{range} we show the optimal sets of the filtering range in
Fourier and corresponding spherical harmonic domain for all PLANCK
frequency channels.

The choice of $\Re=3$ of the initial beam--convolved point source is
to make sure the enhancement $R>5$ for all the PLANCK
channels. Therefore, with the top--hat filtering algorithm, we can
claim that the point sources extracted will have amplitudes above
$3\sigma_{\rm in}$ in the pre-filtered map. Theoretically, we can apply
peak statistics to confirm that the highest point in the
filtered map for the suggested filtering range should be a point
source. The number of peaks above the threshold $\nu_{\rm t}$ per steradian
is 
\begin{eqnarray}
\lefteqn{N_{\max}(\nu_{\rm t})=\frac{\gamma^2}{(2\pi)^{3/2}\theta_*^2}
\nu_{\rm t}
\exp(-\frac{\nu_{\rm t}^2}{2}) +} \nonumber \\
&& \frac{1}{4\pi \sqrt{3} \theta_*^2} 
\left\{ 1-
\Phi\left[\frac{\nu_{\rm t}}{2(1-2\gamma^2/3)^{1/2}}\right]\right\}
\end{eqnarray}\label{eq:peak}
where $\Phi$ is the error function, $\theta_*^2 =
2\sigma_1^2/\sigma_2^2$, and $\gamma=\sigma_1^2/\sigma_0
\sigma_2$ \cite{BE87}. The spectral parameters are defined as 
\begin{equation}
\sigma_i^2=\pi\int {\ell}^{2i+1} C_{\ell}^{\rm tot} d \ell.
\label{eq:spectralpara}
\end{equation}
In order to find the threshold of the highest peak from the
theoretical point of view, we put $A_{\rm f} N_{\max}(\nu_{\rm t})=
1$, where $A_{\rm f}$
is the area of the simulation patch for each channel and $\nu_{\rm t}$ is
the threshold parameter above which there is only one peak. We exclude
 the point source contribution from the input power spectrum
$C_{\ell}^{\rm tot}$ in
Eq.~(\ref{eq:spectralpara}) so that the criterion for point source
extraction should be set above the threshold from the calculation. The
threshold before filtering $\nu_{\rm t}^{\rm i}$ and after filtering
$\nu_{\rm t}^{\rm f}$ for the
corresponding simulation patch of each channel are listed in
Table~\ref{range}. According to the $\nu_{\rm t}^{\rm f}$ values for all channels,
the threshold of the highest peak for the corresponding area after
filtering by the suggested range are less than $5\sigma_{\rm f}$. Hence,
the $5\sigma_{\rm f}$ is an appropriate criterion for point source
extraction.

Figure~\ref{illustration} illustrates an example of
point source enhancement of a realization from the HFI 143 GHz channel. Panel~
(a) shows the simulated map with one added point source of
$\Re=3.21$, i.e., the amplitude of the beam--convolved point source is
$3.21\;\sigma_{\rm in}$.  In panel (b) we display the detailed part of
the beam--convolved point source located at the deepest minimum of the
map. We specifically choose the 143 GHz channel for presentation because
the beam shape carried by the point source is not much deformed by the
pixel noise level. This figure shows the enhancement is related to the
size of FWHM, and the level of the pixel noise against the beam-convolved CMB
plus foregrounds. As shown in Fig.~\ref{powerspectrum1} and
Fig.~\ref{powerspectrum2}, the bending of the power spectrum by the
beam response can be preserved towards high multipole modes as long as
the pixel noise level is low. The less the pixel noise level, the more
towards the high multipole modes the filtering range shifts in order
to include the bending, hence the more low multipole power is excluded
in the filtered $\sigma_{\rm f}$, resulting in higher $R$. For larger FWHM
(e.g. compare the power spectrum of the single point source at 143 GHz
with 353 GHz channel in Fig.~\ref{powerspectrum2}), which nevertheless
bends the total power spectrum, the filtering range has to shift
towards low multipoles to keep the bending part, thus more power is
included in the $\sigma_{\rm f}$, resulting in smaller $R$. Panel (c) is
the filtered map after the filtering ($\kmin,\kmax$) = (42,86)
with enhancement $R=15.5$.

\begin{table}
\begin{center}
 \begin{tabular}{ccccccc}
	 \hline
	 Frequency & $\kmin$ & $\kmax$ & $\lmin$ & $\lmax$ &
	 $\nu_{\rm t}^{\rm i}$ & $\nu_{\rm t}^{\rm f}$ \\ \hline
 857       & 90  & 166  & 2539  & 4673   & $3.45^{*}$ &  4.56 \\ \hline
 545       & 74  & 146  & 2090  & 4111   & $3.73^{*}$ &  4.50 \\ \hline
 353       & 64  & 137  & 1810  & 3859   & $3.85^{*}$ &  4.46 \\ \hline
 217       & 53  & 116  & 1501  & 3269   & $4.01^{*}$ &  4.41 \\ \hline
 143       & 42  & 86   & 1192  & 2427   & $4.11^{*}$ &  4.25 \\ \hline
 100 (HFI) & 68  & 137  & 960   & 1929   & 4.09 &  4.50 \\ \hline
 100 (LFI) & 52  & 116  & 736   & 1634   & 4.31 &  4.39 \\ \hline
 70        & 40  & 93   & 566   & 1311   & 4.43 &  4.29 \\ \hline
 44        & 26  & 68   & 370   & 958    & 3.90 &  4.08 \\ \hline
 30        & 25  & 41   & 356   & 580    & 4.04 &  3.96 \\ \hline
      \end{tabular}
\caption{The optimal filtering range for each PLANCK frequency
channel. The multipole ranges are calculated according to the
corresponding simulation sizes of the maps. As the surface of $R
\equiv R(\kmax,\kmin)$ can vary slightly from realization to
realization due to different deepest minimal value of the realizations
and the effect of cosmic variance, the suggested set of filtering
range can be used as the initial filtering range
$(\kmin,\kmax)^{(0)}$ for iteration schemes. The last two columns
$\nu_{\rm t}^{\rm i}$ and $\nu_{\rm t}^{\rm f}$ are, before and after
filtering, the theoretical values  (in terms of $\sigma$) of the threshold above which there is only one peak
in the corresponding area of the simulated patch. According to
$\nu_{\rm t}^{\rm f}$ from all channels, the choice of $5\sigma_{\rm f}$ is
appropriate for point source extraction. The sign * in the
6th column denotes the values which are calculated from direct
integration of (A1.9) in Bond~\&~Efstathiou(1987).} \label{range}
\end{center}
\end{table}

We set the initial value $\Re=3$ for the estimation of 
the optimal filtering range $(\kmin,\kmax)$. For
$\Re > 3$, the filter with the optimal filtering range will
enhance the index $R$ even more. As tabulated in
Table~\ref{enhancement}, the enhancement for higher frequency channels
can reach above 20, which means that with our proposed method we can
extract point source amplitude much less than $3 \sigma_{\rm in}$ in the
pre-filtered map. We can also estimate  what level of the point source
amplitude at each channel can be detected by the suggested filter. In
Table~\ref{enhancement} we list the amplitudes of the point sources in
terms of $\sigma_{\rm in}$ which can be filtered to reach the criterion
$R=5$ with the suggested range.

\begin{table}
\begin{center}
 \begin{tabular}{cccccc}
	 \hline
	 Frequency & $\Re$   & $R_{\rm TH}$  & $R_{\rm TD98}$  & $R_{\rm MTD98}$ &
	 $\Re_{R=5}$ \\ \hline
 857       & 2.99 & 56.32  & 64.88 & 66.50  &  0.294 \\ \hline
 545       & 3.01 & 36.52  & 41.51 & 41.42  &  0.542  \\ \hline
 353       & 3.05 & 27.46  & 30.71 & 30.61  &  0.867  \\ \hline
 217       & 3.12 & 18.85  & 20.05 & 19.98  &  1.21   \\ \hline
 143       & 3.10 & 14.82  & 15.47 & 15.41  &  1.81   \\ \hline
 100 (HFI) & 3.01 & 11.65  & 12.00 & 11.99  &  1.87   \\ \hline
 100 (LFI) & 2.94 &  6.81  & 6.57  & 6.53   &  2.77   \\ \hline
 70        & 2.91 &  7.36  & 6.82  & 6.68   &  2.41   \\ \hline
 44        & 3.01 &  7.44  & 7.25  & 7.23   &  2.52   \\ \hline
 30        & 3.02 &  6.91  & 6.35  & 6.34   &  2.84   \\ \hline
      \end{tabular}
\caption{Comparison of the enhancing capability for different filters. We
simulate a set of realizations for each frequency channel. The second column
shows the initial amplitude of the beam--convolved point source. The
third column $R_{\rm TH}$ is the result from the top--hat filter. The
fourth column $R_{\rm TD98}$ shows the TD98 filtering and the final
$R_{\rm MTD98}$ is the modified TD98 filtering. The theoretically optimal
TD98 filter gives better enhancement in HFI only when the FWHM of the
beam, dust and noise power spectrum are correctly modelled. The last
column lists the amplitudes of the point sources in terms of the
square root of the variance  of the total map which can be filtered to
reach above $5\sigma_{\rm f}$ by the top--hat filter.}
\label{enhancement}
\end{center}
\end{table}

\section{Comparison with the TD98 filter}
Using the top--hat filter with the optimal filtering range of
$(\kmin,\kmax)$ we can compare its efficiency and accuracy of the
point source extraction from the same realizations with the TD98
filter. The TD98 filter is expressed as 
\begin{equation}
W_k=\frac{B_k}{C_k^{\rm tot}}=\frac{B_k}{B_k^2 C_k+C^{\rm pix}},
\label{eq:TD98}
\end{equation}
where $B_k$ is the beam, $C_k$ is the sum of the power spectrum of the
CMB and foreground components and  $C^{\rm pix}$ is the pixel noise power
spectrum. The shape of
the TD98 filter for each channel is shown with long dash line in
Fig.~\ref{powerspectrum1} and Fig.~\ref{powerspectrum2}, which also
targets the bending of the total power spectrum. Although the TD98 filter is theoretically optimal, the
trouble is that it requires the input of the CMB, and foregrounds power
spectrum and the parameters of the experiments such as the beam size
and shape, and the pixel noise power spectrum. As is claimed in TD98, the
different inputs of cosmological models can have 20 per cent variation in
$R$. In this regard, we also compare the standard (Eq.~(\ref{eq:TD98}))
and the modified TD98 filter, which is mentioned briefly in their
paper. The modified version is that, instead of inserting any
theoretical CMB power spectrum from different cosmological models and
power--law foregrounds into $C_{\ell}^{\rm tot}$ in the filter
($C_k^{\rm tot}$
in the flat sky approximation), we can simply insert $C_{\ell}^{\rm tot}$
from the observed map itself as long as the power spectrum of the CMB
is not severely modified by point sources, which is the case when we
only have one point source in the simulated map. This is to compensate any
fluctuations caused by cosmic variance. The bonus of this is that the
modified TD98 filter does not depend on any cosmological models. In
Table~\ref{enhancement} we show from the top--hat, the TD98 and modified
TD98 filter the enhancement factor of a $3 \sigma_{\rm in}$ point source at the
deepest minimum of the map from a set of realizations. We can see that
for LFI channels, the top--hat filter performs better than both TD98 and
the modified TD98 filters at the {\it worst} situation. We would like
to point out that the TD98 filter is optimal when targeting the
mean $R$, i.e., when a few point sources are located at both
above and beneath the mean level of the map. We therefore list also 
in Table~\ref{enhancementinmean} the enhancement factor $R$ by placing
one point source at a random position of a map and calculate the mean
enhancement factor for a set of realizations. It is shown that the TD98
filter performs better than the top--hat filter, indicating that the
top--hat filtering is $\sim$13-15\% below the optimal extraction of point
sources. This would transfer to roughly 7-10\% loss of point source
extraction compared to the TD98 filter when all the parameters are
known, such as the CMB, dust power spectrum, beam properties, and
the noise level.

\begin{table}
\begin{center}
 \begin{tabular}{cccccc}
	 \hline
	 Frequency  & $R_{\rm TH}$  & $R_{\rm TD98}$  & $R_{\rm MTD98}$  \\ \hline
 857       & 55.91  & 65.29 & 66.81   \\ \hline
 143       & 14.54  & 16.92 & 16.90   \\ \hline
 70        &  7.31  & 8.31  & 8.29    \\ \hline
 30        &  6.94  & 8.24  & 8.21    \\ \hline
      \end{tabular}
\caption{Comparison of the enhancing capability for the filters on
point sources at random positions. Here we show two channels for both
LFI and HFI. We put one point source with $\Re=3$
at a random position of a realization and calculate the mean
enhancement factor $R$ from a set of realizations. The second column
is the result from the top--hat filter with the suggested filtering range in
Table~\ref{range}. The third column $R_{\rm TD98}$ shows the TD98
filtering and the final $R_{\rm MTD98}$ is the modified TD98 filtering.}
\label{enhancementinmean}
\end{center}
\end{table}

To compare from the theoretical point of view the enhancement factor
in details between these two filters for a single point source, we
need the information of the location and amplitude of the point
source. Specifically when we place one point source at the deepest
minimum, this situation favours the top--hat filter, as the top--hat
filter is `designed' for this situation and similar situations such
as other local minima. Also the amplitude of the point source is
crucial in the enhancement factor $R$. Though from
Table~\ref{enhancement} for the HFI channels the $R$ from TD98 filter
is higher, it does become smaller than that from the top--hat filter
when the point source amplitude is significantly smaller than
$3\sigma_{\rm in}$.

\section{Generalization of the top--hat filter}\label{iteration}
The determination of the top--hat filtering range for each channel is
sensitive to the CMB power spectrum model, pixel noise properties,
beam shape and foregrounds contamination. However, for the proposed
ERCSC construction it is absolutely necessary to simplify the method of
the point source detection from the map, using general characteristics
of the `noise' and point sources only. We would like to point out
specifically that the point source filter with adaptive range and a
fixed shape, such as the top--hat filter is a much simpler way to define such kind of filters. As the concept of the top--hat and
the TD98 filter to extract point sources takes advantage of the bending on
the power spectrum by the beam response and the pixel noise level,
they are both useful for the regions near the ecliptic plane in the
flat sky approximation, where the scans are nearly parallel without
crossings (see the fig.1 of Delabrouille, Patanchon \& Audit 2002). For high galactic
latitude scans, however, the crossings of scans complicate the beam
shape configuration. The effective asymmetric beam will manifest
itself in the Fourier domain  which is not isotropic in the Fourier
ring (see the Fig.1 of Chiang~et~al.~2001).  It
is unknown whether the global beam orientation is fixed (parallel) in
the scans of PLANCK. If it is not, it will have degradation effect on
point source extraction from any input of fixed beam function, such as
the TD98 filter and MHW method. 

To tackle this problem and also to relax the condition we set earlier
in Section~\ref{estimation} to determine the filtering range, we would
like to expand the top--hat filter to a more generalized
algorithm. Following Naselsky,~Novikov~\&~Silk~(2002), we can apply the iteration
scheme for point source extraction from the map, using as the initial
step of iterations the suggested $(\lmin,\lmax)$ parameters from
Table~\ref{range}. Without acquiring the exact characteristics of the
experiment such as pixel noise level and beam shape and size, the
initial filtering with the suggested set of $(\lmin, \lmax)$ may
not result in maximal enhancement for that specific realization. We
can, however, always find the highest peak in the filtered map after
the first top--hat filtering. There are two possibilities for the
value of the amplitude of such peak. If we choose the criterion for
the point source detection as $5 \sigma_{\rm f}$, it is likely that the actual
value of the amplitude for the highest peak satisfies this
criterion, from which we can claim we have identified a point source
that can be removed easily from the map. If the highest peak is less
than $5\sigma_{\rm f}$, we can make the second iteration by slightly
fine--tuning the $\lmin$ and $\lmax$ parameter. For the
highest peak $< 5\sigma_{\rm f}$ after the first iteration, if we still cannot
increase its filtered amplitude up to the criterion after fine-tuning,
we can claim that it is not a point source. Of course, for higher
frequency channels of HFI we can set the criterion lower than
$5\sigma_{\rm f}$ as shown in Table~\ref{enhancement}, which means we
can detect point sources with $\Re<3$.

The principal idea is that, with an adaptive filtering range, we can
change $\kmax$ and $\kmin$ step by step to check the
change of $R$ of the highest peak. In each iteration we need to
compare the amplitude of the highest peak (after filtering) with the
criterion for point source detection. For this algorithm, suppose that
for the 100 GHz  channel of HFI we put a point source with $\Re=3$ at
an unknown position in the map and the pixel noise level is twice
 the predicted value. When we apply the filter with the
suggested range, we can detect the highest point, which is less than
$5\sigma_{\rm f}$. By tuning the $\kmax$ with fixed $\kmin$ to
re-filter the original map, we can always find the enhancement $R$ by
going through a bump along $\kmax$. The position of the highest peak in
the filtered map is the same as the initial iteration, but with
different $R$, which indicates our suggested filtering range is not
optimal. Then we fix the $\kmax$ corresponding to the maximum $R$ of the
bump, re-filtering along the $\kmin$ axis to find the $\kmin$
corresponding to the maximal $R$. Through this process we
can always find the new $\kmin$ and $\kmax$ parameter which has
the maximal enhancement $R$.\footnote{In this example the $R$ is now
lower than that listed in Table~\ref{range} as the pixel noise is doubled.}

\section{Discussions and Conclusion}
We have introduced a simple and fast top--hat filter for extraction of
point sources in CMB maps. As the filter cut--off range is
adaptive, we can estimate it with the simulated maps when the
parameters of the PLANCK channels are known. 
    
We would like to emphasise that the main shape of the antenna beam for
the PLANCK mission is close to elliptical due to optical distortions
and telescope designs \cite{burigana2000}. The orientation of the
beam, therefore, is crucial for any methods of point source extraction, which
should be taken into account in order for maximal extraction of point
sources. Our top--hat iteration algorithm, however, does not need this
part of information and it will be even more useful if there is change
of the beam shape due to the degradation effects of the mirrors during
the mission.  

We also perform detailed comparison of the proposed top--hat algorithm
with the TD98 filter. The advantage of our method is that it is very
simple, fast and does not require any detailed information about the
real beam shape, the spectra of CMB and noise, and possible correlations in
the pixel noise. In practice one can take the ready filter with the suggested
range from Table~\ref{range} for each PLANCK frequency channel (if
 desirable) to improve it by the simple and fast algorithm
described in Section~\ref{iteration}. We would like to mention that
the efficiency $R$ for the top--hat filter and the TD98 filter shown
in Table~\ref{enhancement} are practically the same.
  
\section*{Acknowledgments}
This paper was supported in part by Danmarks Grundforskningsfond
through its support for the establishment of the Theoretical
Astrophysics Center and by grants RFBR 17625 and INTAS 97-1192. We
thank Dmitri Novikov for useful discussions and remarks.

\end{document}